\documentclass{sigchi}

\toappear{A demo is available at \url{http://quizcram.herokuapp.com/} \\ Please view it in Chrome.}
\usepackage{balance}  % to better equalize the last page
\usepackage{graphics} % for EPS, load graphicx instead 
\usepackage{txfonts}
\usepackage{times}    % comment if you want LaTeX's default font
\usepackage[pdftex]{hyperref}
\usepackage{color}
\usepackage{textcomp}
\usepackage{booktabs}
\usepackage{ccicons}
\usepackage{todonotes}

\usepackage{paralist}

\makeatletter
\def\url@leostyle{%
  \@ifundefined{selectfont}{\def\UrlFont{\sf}}{\def\UrlFont{\small\bf\ttfamily}}}
\makeatother
\def\pprw{8.5in}
\def\pprh{11in}

\definecolor{linkColor}{RGB}{6,125,233}
\hypersetup{%
  pdftitle={QuizCram: A Quiz-Driven Lecture Viewing Interface},
  pdfauthor={LaTeX},
  pdfkeywords={SIGCHI, proceedings, archival format},
  bookmarksnumbered,
  pdfstartview={FitH},
  colorlinks,
  citecolor=black,
  filecolor=black,
  linkcolor=black,
  urlcolor=linkColor,
  breaklinks=true,
}

\begin{document}

\title{QuizCram: A Quiz-Driven Lecture Viewing Interface}

\numberofauthors{2}
\author{%
  \alignauthor{Geza Kovacs\\
    \affaddr{Stanford University}\\
    \affaddr{Stanford, USA}\\
    \email{geza@cs.stanford.edu}}\\
  \alignauthor{Darren Edge\\
    \affaddr{Microsoft Research Asia}\\
    \affaddr{Beijing, China}\\
    \email{darren.edge@microsoft.com}}\\
}

\maketitle

\begin{abstract}
% Need some motivation here before jumping right into what we built
%In existing MOOC lectures with in-video quizzes, users' video viewing behavior seems to be directed towards solving in-video quizzes.
%Users navigate through the video segments by answering questions.
%QuizCram-format courses can be generated automatically from lectures with in-video quizzes,
% though the format is flexible enough to accommodate multiple questions per video segment.
%We developed it in response to observing that several video seeking behaviors in Coursera center around in-video quizzes.
QuizCram is an interface for navigating lecture videos that uses quizzes to help users determine what they should view. We developed it in response to observing peaks in video seeking behaviors centered around Coursera's in-video quizzes. QuizCram shows users a question to answer, with an associated video segment. Users can use these questions to navigate through video segments, and find video segments they need to review. We also allow users to review using a timeline of previously answered questions and videos. To encourage users to review the material, QuizCram keeps track of their question-answering and video-watching history and schedules sections they likely have not mastered for review. QuizCram-format materials can be generated from existing lectures with in-video quizzes. Our user study comparing QuizCram to in-video quizzes found that users practice answering and reviewing questions more when using QuizCram, and are better able to remember answers to questions they encountered.
\end{abstract}

%\textcolor{red}{Optional section to be included in your final version.}

\category{H.5.2.}{User Interfaces}{Graphical user interfaces (GUI)}{}{}
%  (e.g. HCI)}{Miscellaneous} %\category{See
%  \url{http://acm.org/about/class/1998/} for the full list of ACM
%  classifiers. This section is required.}{}{}

\begin{figure}
\centering
\includegraphics[width=1.0\columnwidth]{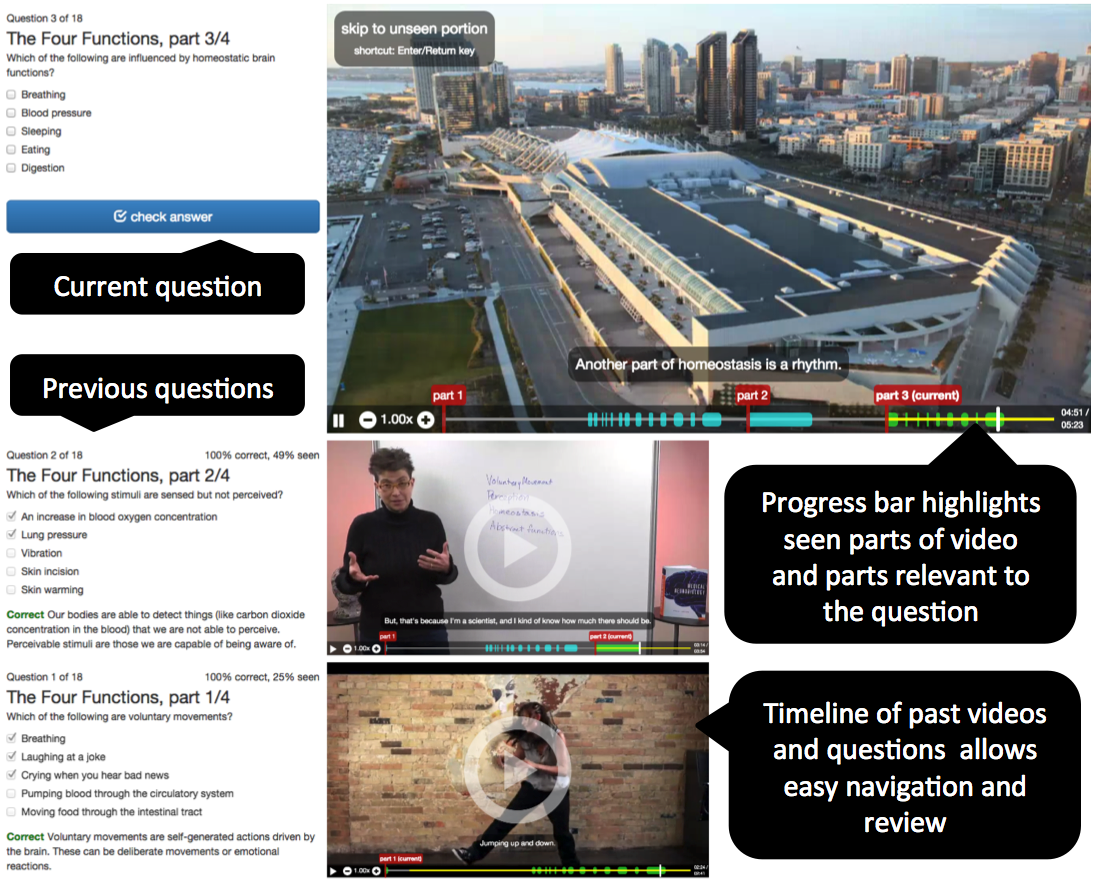}
\caption{The Quizcram interface shows questions on the left, and corresponding video segments on the right. The scrollable timeline displays the past videos and associated questions, to help users review parts they had trouble with.}
\label{fig:timeline}
\end{figure}

% =============================================================================
\section{Introduction}
% =============================================================================
Lectures on platforms such as Coursera use \emph{in-video quizzes} to test learners on material while they watch videos. Although online courses also have problem sets and exams, many learners only watch lectures \cite{anderson2014engaging} \cite{deconstructing}. For these students, in-video quizzes are an important opportunity to test themselves on the material, which is critical for long-term retention \cite{testingeffect}.
%In-video quizzes exploit the \emph{testing effect}, which shows that quizzing improves retention of the viewed material.

While analyzing viewing logs of the Machine Learning course on Coursera, we observed that in-video quizzes play an important role in video navigation. Specifically, we observed that users often seek backward from in-video quizzes to review the preceding section, and forward to in-video quizzes to look at the upcoming question. We also observed that users rarely review lecture videos. Based on these observations, we wished to develop a video viewer that would better support quiz-centric navigation strategies and encourage reviewing.

Our system, Quiz-driven Video Cramming (QuizCram), uses quizzes to help users navigate the course and guide their review process. It includes the following features:

\begin{compactitem}
\item QuizCram shows questions while users watch the video, to serve as a preview of the video content, and to guide their focus towards key concepts.
%\item Our interface shows the question while the user watches the video, so it serves as an advance organizer to prime them towards the key concepts they should focus on %while watching the video.
% \item Our system provides useful feedback in response to an incorrect answer, encourages the user to review the relevant portion of the video, and enforces that users can answer the question on their own before advancing them to the next portion of the video.
% \item To encourage people to review videos, our system keeps track of which video portions users need to review (using a score based on question scores on associated segments, percentage of the segment reviewed, and recency of reviewing), and gives them suggestions of questions and video portions to review once they have watched all the video segments.
\item QuizCram keeps track of which video portions users have already seen, as well as their past performance on questions, in order to suggest which videos and questions the user should review.
%\item QuizCram keeps track of which video portions users have seen, and which they need to review. This is used to give users suggestions of questions and video portions to review. %once they have watched all the video segments.
\item QuizCram facilitates adding questions to videos by allowing questions to depend on multiple video segments rather than just the immediately preceding one. This enables a greater density of questions to be presented in QuizCram.
%\item We facilitate adding questions to videos by allowing questions to depend on multiple video segments rather than just the immediately preceding one. This permits a presentation of broader-reaching concepts and a greater density of questions in the QuizCram format.
%\item We allow questions to be added more easily to videos, by allowing questions depend on video segments other than just the immediately preceding one. This allows for there to potentially be a higher density of questions in the QuizCram format.
\end{compactitem}

We used a user study with a within-subjects design to compare QuizCram to the in-video quiz format. We found that:

% To evaluate the effectiveness of QuizCram for helping users study, we used a within-subjects study design comparing it to an in-video quiz format. Our contributions are:

\begin{compactitem}
% \item QuizCram, a format for viewing lectures in an interactive, question-centric manner, that can be automatically generated from existing videos with in-video quizzes.
%\item Answers to in-video questions are remembered significantly better when users are using QuizCram
\item Users remember answers to in-video questions significantly better when studying using QuizCram.
\item Users practice answering and reviewing questions more often when studying using QuizCram
\item We can improve the recall of particular facts from the video by adding extra questions in QuizCram.
% \item Users are satisfied with QuizCram, and find the interface features for answering questions and reviewing videos to be helpful.
\end{compactitem}

% \pagebreak

%\begin{figure}
\begin{figure*}
\includegraphics[width=2.0 \columnwidth]{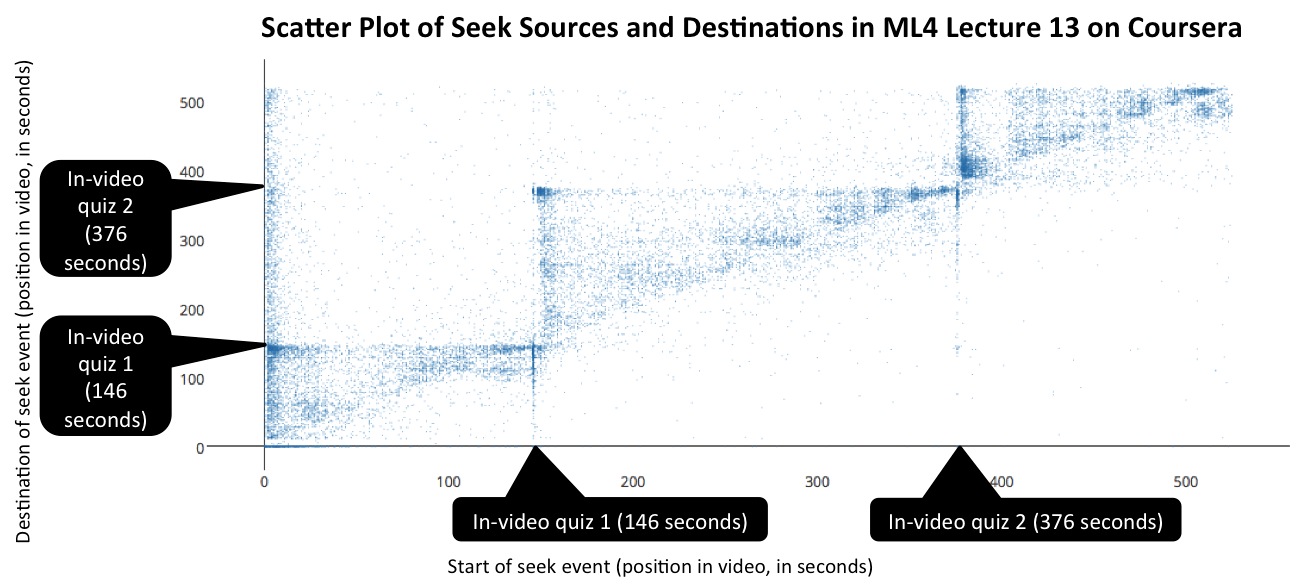}
\caption{Seek sources and destinations in a lecture with 2 in-video quizzes. Each point at (x,y) represents a seek from time x to y. Most seeks do not cross over in-video quizzes. There are many seeks to in-video quizzes from the start of the video, the previous section, and between quizzes.} % Backwards seeks tend to go from in-video quizzes (x=146 and x=376) to the preceding section.
\label{fig:seek-scatterplot}
\end{figure*}
%\end{figure}

\begin{figure}
%\centering
\includegraphics[width=1.0\columnwidth]{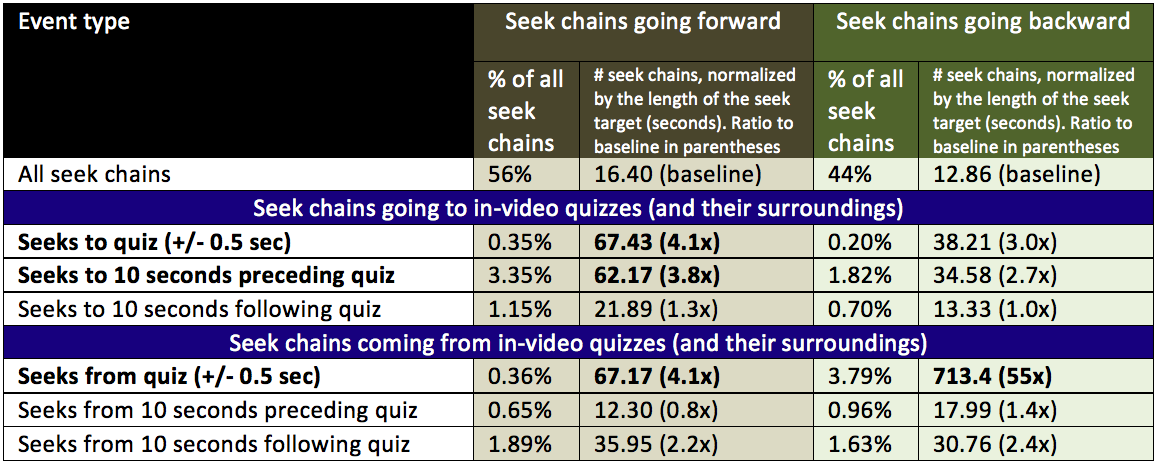}
\caption{Sources and destinations of seek chains in the Machine Learning course on Coursera, which uses in-video quizzes. Users tend to seek backward from in-video quizzes (55x higher than baseline backward-seek rate), and forward to in-video quizzes and the 10 seconds immediately preceding them (4x higher than baseline forward-seek rate)}
\label{fig:seek-sources-and-destinations-short}
\end{figure}

\begin{figure}[h]
%\centering
\includegraphics[width=1.0\columnwidth]{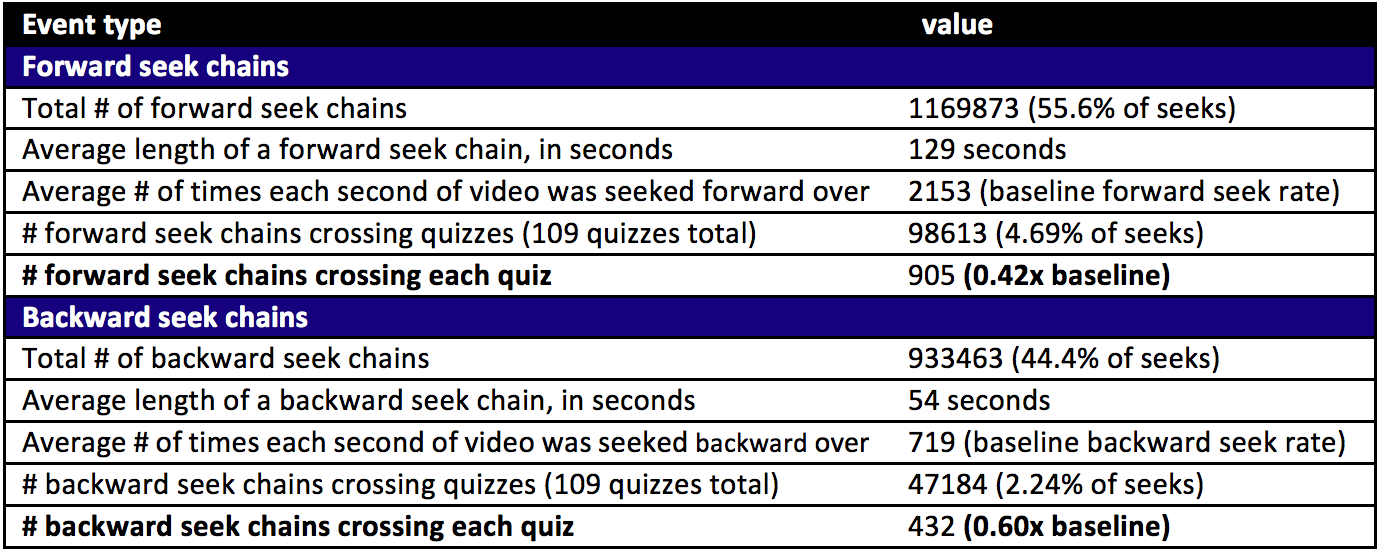}
\caption{Portions of the video that are skipped over by seek chains in the Machine Learning course on Coursera. Users do not tend to seek forward across in-video quizzes.}
\label{fig:table-of-seeks-short}
\end{figure}

\section{Motivation: Coursera's In-Video Quizzes}

% During our analysis of viewing logs for the Machine Learning course on Coursera, we observed that users' navigation behavior is heavily influenced by in-video quizzes.

% This work was motivated by seeking activity around in-video quizzes we observed while analyzing Coursera's viewing logs for the Machine Learning course. We observed that there are large peaks in seeking activity around in-video quizzes, which is likely due to users previewing the questions and trying to find answers to them. % The dataset we analyzed was the viewing logs for the fourth offering of the Machine Learning course, a MOOC which has been topic of much previous research \cite{anderson2014engaging}. % This course contains 109 in-video quizzes across 19.5 hours of video.

This work was motivated by interesting patterns of seeking activity around in-video quizzes which we uncovered while analyzing viewing logs of Coursera's Machine Learning course. We observed that there are large peaks in seeking activity around in-video quizzes, which is likely due to users previewing the questions and trying to find answers to them. 

% This work is envisioned as an enhancement upon the in-video quizzes used in Coursera. Hence, our first step was to analyze user interactions with in-video quizzes in Coursera's viewing logs. The course we analyzed was the .

Since users may seek several times while trying to reach their target, our analysis groups together seek events that occur within 5 seconds of each other into a \textit{seek chain}, so we can better observe users' intended seek targets. Details on this methodology can be found in the supplement.

% Our analyses of users' in-video navigation focuses on \textit{seek chains}, which are the result of grouping together seek events that occur

There are many backward seeks starting from in-video quizzes. As shown in \autoref{fig:seek-sources-and-destinations-short}, 8.6\% of all backward seek chains (or 3.8\% of total seek chains) start from in-video quizzes -- which is 55x more seeking per in-video quiz than we'd expect from a second of video in the course. This peak in backward-seeking from in-video quizzes is likely due to users searching for answers in the preceding section.

We also observe that there are many forward seeks that end up at or immediately preceding the quiz. As shown in \autoref{fig:seek-sources-and-destinations-short}, 6.6\% of all forward seek chains (or 3.7\% of total seek chains) end up either at the in-video quiz or within 10 seconds preceding it. These forward seeks are likely generated by users attempting to view the in-video quiz -- as Coursera's interface does not provide an option to jump directly to in-video quizzes, users must seek to directly before the in-video quiz in order to view it.

%  -- users do not tend to skip forward or backward across in-video quizzes.
Most seek chains (93\%) do not cross in-video quiz boundaries. As shown in \autoref{fig:table-of-seeks-short}, users are 0.4x less likely to skip forward across an in-video quiz, than across a second of video. \autoref{fig:seek-scatterplot} visualizes seek sources and destinations in a single lecture video with 2 in-video quizzes: there are many forward seeks to quizzes, and backward seeks from quizzes.

Users also rarely rewatch lecture videos: only 11\% of users who finished watching a lecture will ever open it again.

Based on these findings, we aimed to develop a video viewer that would better support quiz-centric navigation strategies and encourage reviewing.

% Further details and methodology for this logs analysis work can be found in the supplement.

% . There is much seeking to and from in-video quizzes -- 6.5\% of forward seeks go to in-video quizzes or the 10 seconds preceding them, even though this is ,  destination of seeks users are 3.5 times more likely to seek to the 10 seconds before a quiz than to other segments.  Users often skip forward to quizzes, or from one quiz to the next, as shown in \autoref{fig:seek-scatterplot}.

% =============================================================================
\section{Related Work}
% =============================================================================
%We designed QuizCram's features based on the following findings from the education literature:
%We designed QuizCram's features based on the testing
% In addition to our findings about peaks in seeking around in-video quizzes (see supplement), we also designed QuizCram's features based on the following findings from the education literature:%, which are also exploited by many other systems.

\subsection{Testing and Pre-Testing Effects}

The \emph{testing effect} shows that repeated testing combined with fast, informative feedback helps students remember material \cite{testingeffect}. QuizCram's emphasis on answering and reviewing questions is designed to exploit this effect.

% The testing effect finds that repeated testing combined with fast, informative feedback helps students remember material \cite{testingeffect}. In-video quiz systems are based on this principle: by testing the user on the video contents that were just viewed, they help students remember the material \cite{guidingquestions}. QuizCram's question-directed studying approach is also designed to exploit the testing effect.

%The \emph{pre-testing effect} shows that asking users to try answering a question before they actually study the material enhances long-term retention \cite{pretesting}. QuizCram leverages the pre-testing effect by allowing users to preview the question before watching the associated video.

The \emph{pre-testing effect} shows that having users try answering a question before they actually study the material enhances long-term retention \cite{pretesting}. QuizCram exploits the pre-testing effect by allowing users to preview the question before watching the associated video.

\subsection{Spaced repetition}

Spaced repetition is a technique designed to help learners retain information by having them review  items at regular intervals \cite{karpicke2011spaced}. A class of applications that exploit this are flashcards, which split information into independent chunks that are scheduled for review based on factors such as mastery and recency of review. There have been a number of algorithms and models designed for optimizing learners' retention of the material via spaced repetition \cite{optimalschedule} \cite{memreflex}. However, they tend to be designed for flashcard-like content, such as isolated facts or vocabulary, rather than lecture videos. % Flashcards can also have associated multimedia, such as video clips.

A key difference between flashcard-like content and lecture videos is that lecture videos are typically presented in sequence, and a given video may build upon concepts introduced in a previous video. Additionally, there are differences in the costs of testing and reviewing. With flashcards, both testing and reviewing can be done in seconds. In contrast, the cost of reviewing a video is much greater than the cost of testing -- we can test a user's knowledge of a video segment with a question that takes seconds to answer, but viewing a video may require several minutes. These additional constraints are reflected in QuizCram's modified scheduling algorithm that takes into account the order of videos, as well as its increased emphasis on testing via questions.

% Spaced repetition is a technique used by flashcards to help learners retain information by having them review items at regular intervals \cite{karpicke2011spaced}. Similar to flashcards, our system also schedules questions for review based on our model of the user's mastery and recency of review. % A key difference is that lecture videos build on each other, so we must schedule earlier videos first.

% Spaced repetition is a technique designed to help learners retain information by having them review  items at regular intervals \cite{karpicke2011spaced}. A class of applications that exploit this are flashcards, where information is split into independent chunks that are scheduled for review based on factors such as mastery and recency of review. Similar to flashcards, our system also schedules items for review according to mastery and recency of review.

% Similar to flashcards, our system also schedules items for review according to mastery and recency of review. One key difference is that lecture videos build on each other, so this is an additional constraint for scheduling: the user needs to have covered the previous videos. % Another key difference is the cost of review: a user memorizing vocabulary using flashcards only needs to spend a few seconds on each flashcard, while answering a question or reviewing a video clip takes an order of magnitude more time. Hence, the user will make fewer review passes through the video content than they would with vocabulary flashcards.

\subsection{Advance Organizers}

Advance organizers are information presented prior to learning, that help the learner process the material that is about to be presented  \cite{advanceorganizers}. QuizCram's questions can be thought of as an advance organizer for the video segment -- the question provides a preview of the content that is to be covered in the video. % Video Digests is a system that creates such summaries about videos, and uses them as an advance organizer and navigational guide for video lectures \cite{videodigests}.

\subsection{Interfaces for Navigating Lecture Videos}

Video Digests is a system that uses textual summaries of video clips to help users navigate through the video \cite{videodigests}. LectureScape uses other users' aggregated viewing logs to help identify points of interest in the video \cite{lecturescape}. Panopticon uses a visual display of all video segments to help users find segments of interest \cite{panopticon}. Similar to these systems, QuizCram aims to help users navigate through lecture videos. However, rather than relying on external annotations, QuizCram instead uses questions extracted from existing in-video quizzes as a navigational aid. % This has the advantage over  %  Unlike Video Digests and LectureScape, QuizCram does not require 

% QuizCram similarly breaks videos into segments associated with an advance organizer, but we use existing in-video questions to summarize the clips. Both QuizCram and advance organizers use a short task (reading some text or answering a question) to help the user decide whether to invest several minutes in viewing/reviewing the video.

% Advance organizers are information presented prior to learning, that help the learner process the material that is about to be presented  \cite{advanceorganizers}. An example of an advance organizer for lecture video content would a summary of the video content that is to be watched. Video Digests is a system that creates such summaries about videos, and uses them as an advance organizer and navigational guide for video lectures \cite{videodigests}. Our system follows a similar strategy of breaking the video into segments associated with an advance organizer, but we instead use a question as an advance organizer that summarizes the clip to the user before they start watching it.

% =============================================================================

% =============================================================================

\pagebreak

\section{System Design Process}

Based on our observations that users tend to engage with in-video quizzes but rarely ever revisit MOOC lecture content (see supplement), as well as the importance of testing and review for retention, our goal was to build a system that would test users' knowledge of lecture materials and encourage them to review materials using spaced repetition.

% \subsection{Design Process}

Our initial design was to treat video segments as flashcards, and schedule them using a spaced repetition algorithm. By associating each video segment with a question, we could easily test users's knowledge of each segment. However, scheduling videos with a standard spaced-repetition algorithm would often result in the user being asked to review older material before they completed all of the video segments, which we found that users were unaccustomed to. Hence, we also enabled users to freely review videos on their own, and only started scheduling older videos for review once they had attempted an initial pass through the videos. %and only start scheduling older videos for review based on spaced repetition, once they had made an initial pass through the videos.

\section{QuizCram Interface Features}

% QuizCram's interface shows users a question to review, with an associated video segment, as shown in \autoref{fig:timeline}. It also shows a scrollable timeline of previously answered questions and associated video segments below the current question. Questions are first scheduled in order, then once the user has made an initial pass, questions are selected for review algorithmically, based on historic correctness of responses, percentage of associated video that has been watched, and the recency of review. We also use the video progressbar to indicate the section of the video that is relevant to the current question, and portions of the video that the user has previously seen.

\begin{figure}
\centering
\includegraphics[width=1.0\columnwidth]{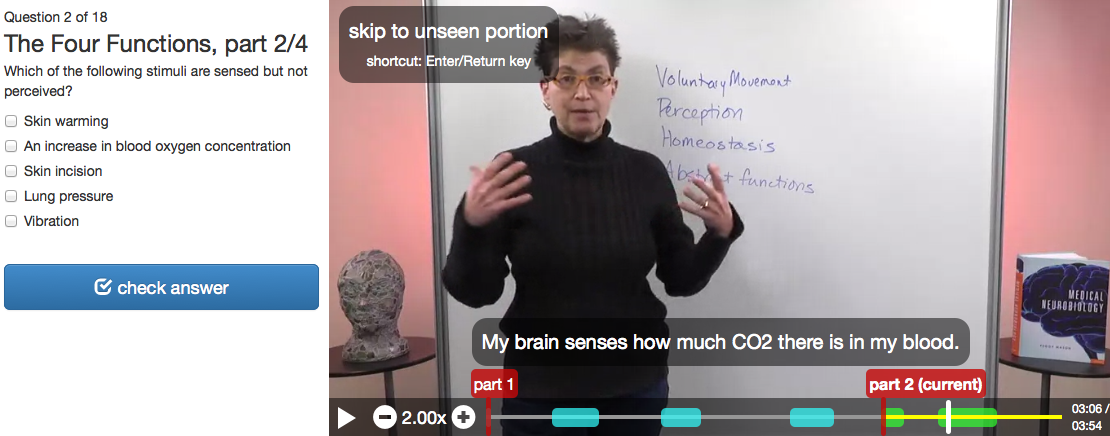}
\caption{The QuizCram interface, showing the focus question on the left, and the associated video on the right. The progress bar highlights the relevant portion of the video in yellow. Segments that have already been watched are highlighted in blue (segments from previous parts) and green (segments from current part).}
%\caption{The QuizCram interface, showing the current video. The focus question is on left, and the associated video is on the right. The progressbar highlights the relevant portion of the video in yellow. Already-watched segments of previous sections is in blue, already-watched segments of the current part are in green. Because we are currently watching a section we have already viewed, an option to skip to the unseen portion is shown.}
\label{fig:singlevideo}
\end{figure}

QuizCram's interface displays a question and associated video segment, as shown in \autoref{fig:singlevideo}. It also shows a timeline of previous questions below the current question, as shown in \autoref{fig:timeline}. Once the user has made an initial pass through the questions, we suggest questions that they should review, based on past performance. We use the video progress bar to indicate the section of the video that is relevant to the current question, and portions that the user has previously seen. Existing courses with in-video quizzes can easily be transformed into the QuizCram format. 
%Existing courses with in-video quizzes, such as MOOCs on Coursera, can easily be transformed into the QuizCram format. 

% An existing course with in-video quizzes, such as MOOCs on Coursera, can be automatically transformed into the QuizCram format. This results in each video segment having one associated question. However, unlike in-video quizzes, the QuizCram format can also have multiple questions associated with a single video segment.

% =============================================================================
\subsection{Question-Directed Video Viewing}
% =============================================================================
Each video section is associated with a question. We can extract these question-video pairs automatically from existing videos with in-video quizzes, by associating the in-video quiz section with the immediately preceding video segment. For video segments that do not have an associated in-video quiz, we can either automatically insert a generic ``How well did you understand this video'' question, or manually write a new question.

The question is designed to help users decide whether they should watch the video. If the user knows the answer, they can answer the question and move to the next section. For users who do not  know the answer, reading the question provides a preview of the key points they will see in the video.

\subsection{Timeline of Previous Questions and Videos}

The \emph{timeline} feature is designed to encourage review by making it easy to refer back to previously answered questions and video segments. Whenever a question is correctly answered, we insert the next question and associated video segment at the top of the interface, and push the existing questions down. This results in a scrollable visual history of previously answered questions, as shown in \autoref{fig:timeline}. The timeline displays the question, its answer, and a miniaturized version of the video which can be clicked to enlarge it to full size and play it. The miniaturized video displays the frame the user left off at, so it serves both as a visual summary, and also allows users to easily resume watching previous videos.% where they left off.

%By organizing the list of previous video segments according to the associated question that users answered, this allows users to scan video segments with a more salient summary than just the title. Question-based video navigation also allows users to search at a finer granularity, as questions refer to a specific subsection of the video, while the title refers only to the entire video contents.

% Although QuizCram focuses the user's attention towards the current question and associated video segment, we also wish to make it easy to refer back to the previously answered questions and video segments. Whenever a question is correctly answered, we insert the next question and associated video segment at the top of the interface, and push the existing questions down. This results in a scrollable visual history of the previously answered questions and videos which we call the \emph{timeline}, shown in \autoref{fig:figure3}. The timeline displays the question and its answer and a miniaturized version of the video which can be clicked to enlarge it to full size and play it. The miniaturized video displays the frame the user left off at, so it serves both as a visual summary, and also allows users to easily resume viewing progress of previous videos. % We also show the historic correctness of the user's answers to that question, and percentage of the video they have watched, to help users identify questions they had trouble with and videos they did not fully watch.

The timeline gives users the option to use a more self-directed reviewing strategy, in contrast to the flashcard-style reviewing that our question scheduling algorithm encourages. By organizing the list of previous video segments according to the associated question that users answered, this allows users to scan video segments with a more salient summary than just the video title. Furthermore, re-reading the previously answered questions can help trigger the users' memory of the associated video clips

%The timeline gives users the option to use a more self-directed reviewing strategy, in contrast to the flashcard-style reviewing that our question scheduling algorithm encourages. By organizing the list of previous video segments according to the associated question that users answered, this allows users to skim through the video list at a higher granularity, as questions refer to a specific subsection of the video, while the title refers only to the entire video contents. Furthermore, re-reading the previously answered questions can help trigger the users' memory of the associated video clips. %, giving learners another retrieval opportunity to solidify their memory of the video contents.

% The timeline gives users the option to use a more self-directed reviewing strategy, in contrast to the flashcard-style reviewing that our question scheduling algorithm encourages. By organizing the list of previous video segments according to the associated question that users answered, this allows users to scan video segments with a more salient summary than just the title. Question-based video navigation also allows users to skim through the video list at a higher granularity, as questions refer to a specific subsection of the video, while the title refers only to the entire video contents. % Furthermore, re-reading the previously answered questions helps trigger the users' memory of the associated clip, giving learners another retrieval opportunity to solidify their memory of the video contents.

\subsection{Scheduling Questions and Video Sections for Review}

We want users to spend their study time focusing on material that they have not yet mastered. Hence, we assign each question a \emph{mastery score}, which represents how well the user currently knows the material, and show users the questions for which they have low mastery score. The question's mastery score is based on the following 3 factors:

%\begin{compactitem}
\begin{itemize}
\item \emph{Past performance on question}: This element of the score encourages users to review questions they answered incorrectly. Each time a user tries answering a question, we give them a score equal to the fraction of checkboxes they correctly checked (the questions used in our study were all multiple-check questions). We then take a weighted-mean of historic scores, weighing recent answers more heavily. %all historic scores, with each newer score assigned 2 times more weight than the previous score (so more recent performance is weighted more heavily). %For those video segments that have no associated question, we obtain this score by asking users to rate ``How well did you understand this video?''. If the user has never answered the question before, this has a default score of 0.
\item \emph{Fraction of associated video segment watched}: This element of the score encourages users to view video segments they have not seen. For each second of video, we keep track of whether the user has ever seen it. This score is the fraction of the video segment that has been seen. % number of seconds watched in the question's video segment, divided by the total length of that video segment.
%\item \emph{Recency of review}:  This element of the score ensures that users review old questions, but are not shown same questions repeatedly. For simplicity, we use a score that is inversely proportional to how recently the question was last answered. Ideally, one would instead use overdueness of items from a more advanced spaced-repetition algorithm like MemReflex \cite{memreflex}.
\item \emph{Recency of review}:  This element of the score ensures that users review old questions, but are not shown same questions repeatedly. For simplicity, we use a score that is inversely proportional to how recently the question was last answered. Ideally, one would instead use a more advanced spaced-repetition algorithm like MemReflex \cite{memreflex}. %It is equal to 1 / number of questions elapsed since this question was last seen by the user. If the question has never been seen, this has a default score of 0.
%\end{compactitem}
\end{itemize}

% and show users questions from areas where their mastery scores are low.

% We want users to spend their study time focusing on material that they have not yet mastered. Hence, we assign each question a \emph{mastery score}, which represents how well the user currently knows the material, and show users the questions for which they have low mastery score. The mastery score is a weighted sum based on the user's past performance on the question, the fraction of the associated video segment they have watched, and the recency of review.

% We want users to spend their study time focusing on material that they have not yet mastered. Hence, we assign each question a \emph{mastery score}, which represents how well the user currently knows the material,  and show users questions from areas where their mastery scores are low. The mastery score is a weighted sum based on the user's past performance on the question, the fraction of the associated video segment they have watched, and the recency of review.

Once the user has seen all the questions in the unit, QuizCram encourages them to review questions and sections for which they have low mastery scores, by showing them at the top of the video timeline.

\subsection{Directing Attention to Unseen Parts of Videos}

To help users review videos and resume where they left off, QuizCram keeps track of which parts have been watched. It highlights on the progress bar the portions that have already been seen. If the user is viewing a section they have already watched, they can skip to the unseen portion by clicking a button, as shown in \autoref{fig:singlevideo}. This technique for visualizing the viewing history has previously been shown in the literature \cite{socialnavigation} \cite{lecturescape}, though our system adds the novel feature of allowing users to skip to the next unseen portion. % Similar techniques for visualizing the user's video viewing history have been presented in the literature \cite{socialnavigation} \cite{lecturescape}, %though our system adds the novel feature of allowing users to skip to the next unseen portion.

% =============================================================================
\section{Evaluation}
\begin{figure}
\centering
\includegraphics[width=1.0\columnwidth]{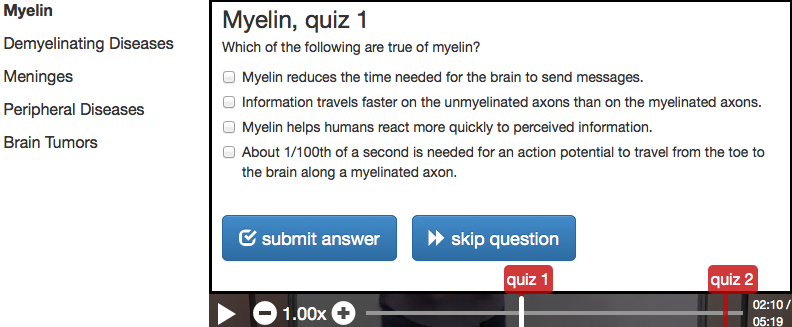}
\caption{The in-video quiz format that served as our baseline. Locations of quizzes are indicated in red on the progress bar.}
\label{fig:invideo-interface}
\end{figure}

% =============================================================================
Our study used a within-subjects design to compare users' studying behavior with QuizCram against an in-video quiz interface that mimcs the format used on Coursera, as shown in \autoref{fig:invideo-interface}. We used the videos, in-video quizzes, and unit exam from the Neurobiology course on Coursera. We wished to answer the questions:

\begin{compactitem}
\item Does QuizCram help users better remember answers to the original in-video questions?
\item Does QuizCram help users score higher on exams?
\item Do users find QuizCram helpful for studying videos?
\item How do users interact with questions and videos when using QuizCram?
%\item Can we improve recall of particular facts from the video by adding extra questions with QuizCram?
%\item Do users using QuizCram answer questions more?
%\item Do users using QuizCram re-answer questions more?
\end{compactitem}

%\subsection{Study Design}

%The study was a within-subjects design, where each learner used QuizCram and an in-video quiz viewer interface to study a set of videos. They were asked to provide qualitative feedback immediately after viewing, and were tested on the material they studied a day later.

\subsection{Participants}

We recruited 18  students by posting on university mailing lists. 12 were female, 6 male. Their average age was 21.7 ($\sigma$=4.91, min=18, max=37). All had native-level English proficiency. None had prior exposure to neuroscience.  They received \$60 for participating. %in the 2-hour online study. % 9 participants reported having previous experience with MOOCs, and of these 6 had experience with Coursera.

\subsection{Materials}

% The course materials -- videos, in-video quizzes, and unit exams --- were the first and second halves of Unit 1 of an existing Neurobiology course on Coursera. We generated the initial QuizCram materials directly from the course. However, because we felt the question-to-video ratio in the original videos (9 questions for each 25-minute segment) was lower than what would be optimal for QuizCram, and because we wished to see the effects of inserting additional questions on recall of associated facts, we wrote additional questions for the QuizCram condition to double the total number of questions.

% We doubled the number of questions shown in the QuizCram condition by adding additional questions in the same style and format as the original multiple-checkbox in-video questions. We chose our questions carefully such that the answers were clearly stated in the video, but they would not ask the same facts that were tested on the unit exam or original in-video questions.

The videos, in-video quizzes, and unit exams were from Unit 1 of the Neurobiology course on Coursera. There were 9 questions and 5 videos in each 25-minute section. We generated the initial QuizCram materials directly from the course.

% During pilot studies, we noticed that large segments of the videos were not covered by 
% Because large segments of the video were not covered by any, 
% . In a pilot study -- however, 
% Because the original in-video quiz materials did not . In a pilot study, we had . We envision that if material is designed specifically for the .

% Furthermore, we believe that the QuizCram format is best-suited towards a more question-heavy viewing experience than in-video quizzes currently provide. 

Not all of the segments of the lecture videos had in-video quizzes immediately following them. For such segments, QuizCram would normally show a generic ``How well did you understand this video'' question as the focus question. However, in pilot studies, users indicated that they found these self-assessment questions less helpful than regular questions, as they did not provide a preview of what the section would be about. Furthermore, we believe that the QuizCram format is best-suited towards a more question-heavy viewing experience than in-video quizzes currently provide. Hence, to simulate what content that was designed for the QuizCram format would look like, we added our own extra questions for video segments which lacked associated in-video quizzes. This doubled the total number of questions per section in the QuizCram condition. The extra questions were in the same multiple-checkbox format as the original questions. We made sure that the extra questions did not depend on the same facts as the unit exam or original in-video quizzes, to ensure that they would not help users learn the other material by giving them an extra testing opportunity. %did not cover the same material as the unit exam or original in-video quizzes, to ensure that they would not help users learn the other material by giving them an extra testing opportunity.

% To see whether we could improve the recall of particular facts by adding questions, we added extra questions to the QuizCram condition to double the total number of questions. The extra questions were in the same multiple-checkbox format as the original questions. We made sure that they did not depend on the same facts as the unit exam or original in-video quizzes.

We also wrote a set of free-response questions, with one corresponding to each of the extra questions. We used these free-response questions to test whether users had learned the material tested by in-video questions well enough to recall it (rather than recognizing it).

% We also wrote a set of free-response questions, one corresponding to each of our extra multiple-checkbox questions. For example, the question ``Which of the following are true of astrocytes?'' followed by 3 true options and 2 false options would be transformed into the free-response question ``List 3 facts about astrocytes''. We used these free-response questions to test whether users had actually learned the material tested by our new questions well enough to recall it, as opposed to simply learning to recognize the answers when presented in multiple-checkbox format.

\subsection{Procedure}

The study was conducted online over 2 days. Before users started the study, we informed them that they would be given 2 sets of videos, they would study them for 40 minutes apiece, and they would be given an exam the next day. We did not tell them about the content of the exam in advance. % Our server logged interactions as they were using the tools. %, because we had observed in pilot studies that some users would 

% The study was conducted online over 2 days, with a 90-minute study session on the first day, and a 30-minute test session on the second day. Before users started the study, we informed them that they would be given 2 sets of videos, they should study them for 40 minutes apiece, and they would be given an exam the next day. There was a 10-minute break between the study sessions. We did not tell them about the content of the exam in advance. Our server logged interactions as they were using each of the tools. %, because we had observed in pilot studies that some users would 

%The study was conducted online over 2 days. On day 1, users were given 40 minutes to study the first section with one tool. Then, they used the other tool to study the second section for 40 minutes, and filled out a survey (order of tools was randomized). On day 2, users took the following exams:

On day 1, users studied the first section with one tool for 40 minutes, and answered a survey about the tool. Then, they studied the second section with the other tool for 40 minutes, and answered a survey about the tool. The order of tools was randomized. % Our server logged interactions with each of the tools.

% On day 2, users took the following exams:

% The study was conducted online over 2 days.

% On day 1, users used one tool to watch the first half of Unit 1 (5 videos with a total length of 23 minutes). They were told after 40 minutes to fill out a survey about the tool. Then, they used the other tool to watch the second half of Unit 1 (5 videos with a total length of 25 minutes), and filled out the survey after 40 minutes of watching. The order of tools was randomized. % Before they started studying, we told them they would be given an exam the following day about the material. However, we did not tell them the specific format or contents of the exam in advance.

On day 2, users took the following exams:

\begin{compactenum}
\item Extra free-response questions %, both halves
\item Original in-video questions from Coursera%, both halves
\item Original unit exam from Coursera%, both halves
\item Extra multiple-checkbox questions %, both halves
\end{compactenum}

%Parts 2-4 of the exam were automatically graded, giving each question a score equal to the fraction of checkboxes correctly checked. % The free-response questions, which were of the general form ``List N examples of X'' or ``State N facts about X'', were graded by first marking each example provided by students as correct or incorrect. % Examples and facts did not need to be the same ones provided in the video -- so long as they were appropriate to the question and correct, they were marked as correct.
% Then, we scored each response via the formula:
% Parts 2-4 of the exam were automatically graded, assigning scores equal to the fraction of checkboxes correctly checked.
% The free-response questions, which were of the form ``List N examples of X'' or ``List N facts about X'', were scored as:
%Parts 2-4 of the exam were automatically graded, assigning scores equal to the fraction of checkboxes correctly checked. The free-response questions were graded by an independent grader.
% Parts 2-4 were automatically graded, and free-response questions were graded blindly. % according to the formula: %by an independent grader.
% Free-response questions were graded by an independent rater.

Parts 2-4 were automatically graded. Free-response questions were graded blindly according to the formula:

\vspace{-4mm}

\[ \frac{\# correct\ examples\ given}{Maximum(\# examples\ requested,\ \# examples\ given)} \]

% Thus, if a question requests 2 examples, giving 1 correct example gets a score of 1/2, giving 2 correct examples and 1 incorrect example gets a score of 2/3, etc. The overall score for the free-response exam was the mean of these scores.

% We chose this design of having users watch both sets of videos before taking any exams, as opposed to having them take an exam after they finished studying each section, because this way the user does not know that the exam includes the in-video questions, so this does not influence their study behavior. In previous versions of this study we had observed that if users know they will be tested on in-video questions, either by taking the exam or if we told them, they will explicitly study them and will have near-perfect scores on those parts of the exam, but not the rest of the exam. We instead chose our current study design so that we can observe the tool's effect on in-video question retention in natural study contexts where the user knows nothing about the exam.

\section{Results}

\subsection{Exam Results}

\begin{figure}
\centering
\includegraphics[width=1.0\columnwidth]{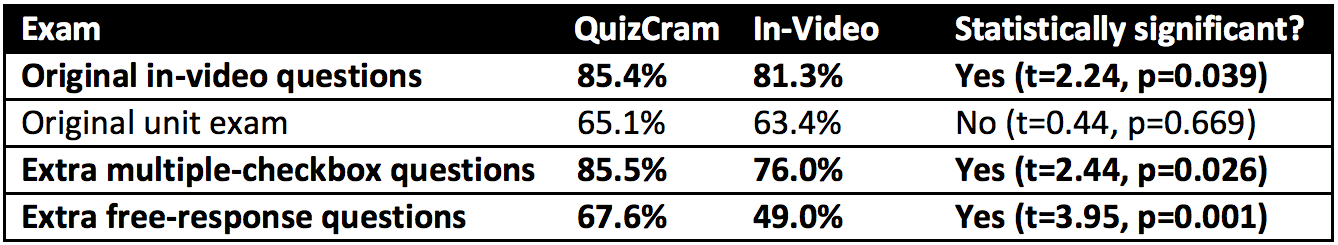}
\caption{Average exam scores for each condition}
\label{fig:exam-results}
\end{figure}

Exam results are shown in \autoref{fig:exam-results}.  QuizCram users performed significantly better on the original in-video questions, which had been shown in both conditions. They also performed better at both types of extra questions. Thus, QuizCram improves retention of the original in-video questions, and we can use added questions to improve retention of particular facts from the video. However, there was no significant improvement in scores on the original unit exam. % from Coursera.
% We attribute this improvement to QuizCram's increased focus on questions.
% There was no significant difference in unit exam scores.

\subsection{Survey Results}

\begin{figure}
\centering
\includegraphics[width=1.0\columnwidth]{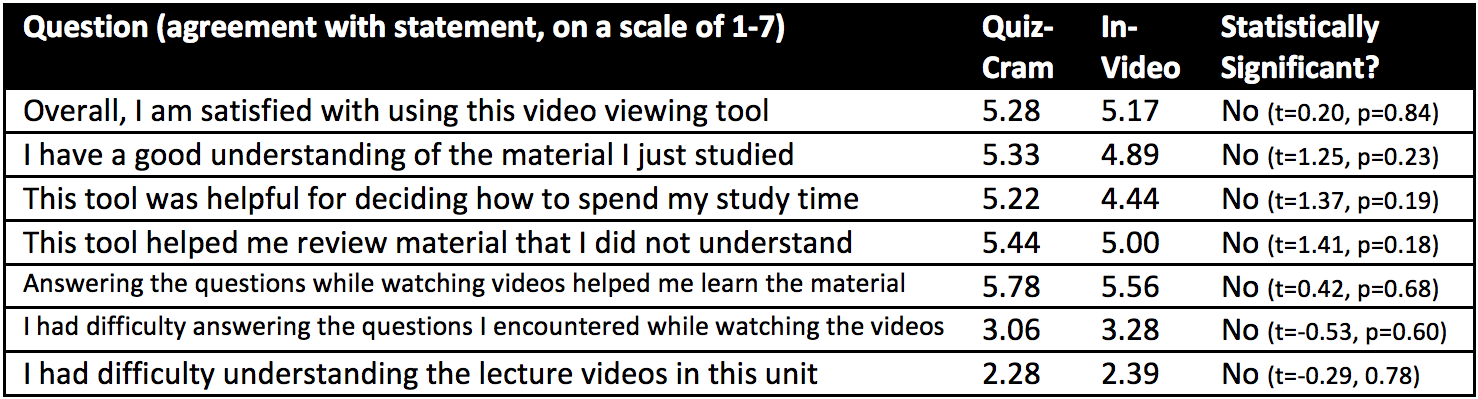}
\caption{Survey responses showed slight preferences in favor of QuizCram, but they were not statistically significant}
\label{fig:survey-results}
\end{figure}

Survey responses after using each tool are shown in \autoref{fig:survey-results}. 61\% said would prefer to use QuizCram if they wanted to remember material long-term or were preparing for an exam. These improvements were not statistically significant.

%When users were asked to rate satisfaction with the tool on a scale of 1 to 7, the average rating was 5.28 for QuizCram, and 5.17 for in-video quizzes.  61\% said would prefer QuizCram if they wanted to remember material long-term or were preparing for an exam. These improvements were not significant.

%Survey feedback showed that users liked QuizCram's question-based timeline of videos, and thought it was helpful for reviewing. However, some users disliked the visual complexity of the interface, and thought that the prominent display of questions distracted them from watching the video.

%Survey feedback showed that users liked QuizCram's question-based timeline of videos, and thought it was helpful for reviewing.

%Survey feedback showed that users thought QuizCram's features were helpful for reviewing:

Survey feedback showed that users thought QuizCram's focus questions helpful for reviewing videos:

\textit{I liked that it picked out the key information I should retain by asking me questions. It helped me decide what to focus on as I watched the video. The chunks were very manageable as well. I liked how it was broken up.}

%\textit{It was easy to review the questions and quiz myself multiple times, and ultimately, the repetition aided my understanding of the material.}

%\textit{The video viewing tool gave you the opportunity to learn the material a second time by presenting you the question again out of order (and without the answers). I found that I had to rely on my own recall to answer the questions a second time, and learned the material better.}

%However, some users disliked the visual complexity of the interface, and thought that the prominent display of questions distracted them from watching the video.

However, some users thought that the prominent display of questions distracted them from watching the video.

\textit{I did not like the fact that you could answer questions while the video was playing. It made me more focused on answering the questions rather than watching and learning the material.}

\begin{figure}
\centering
\includegraphics[width=1.0\columnwidth]{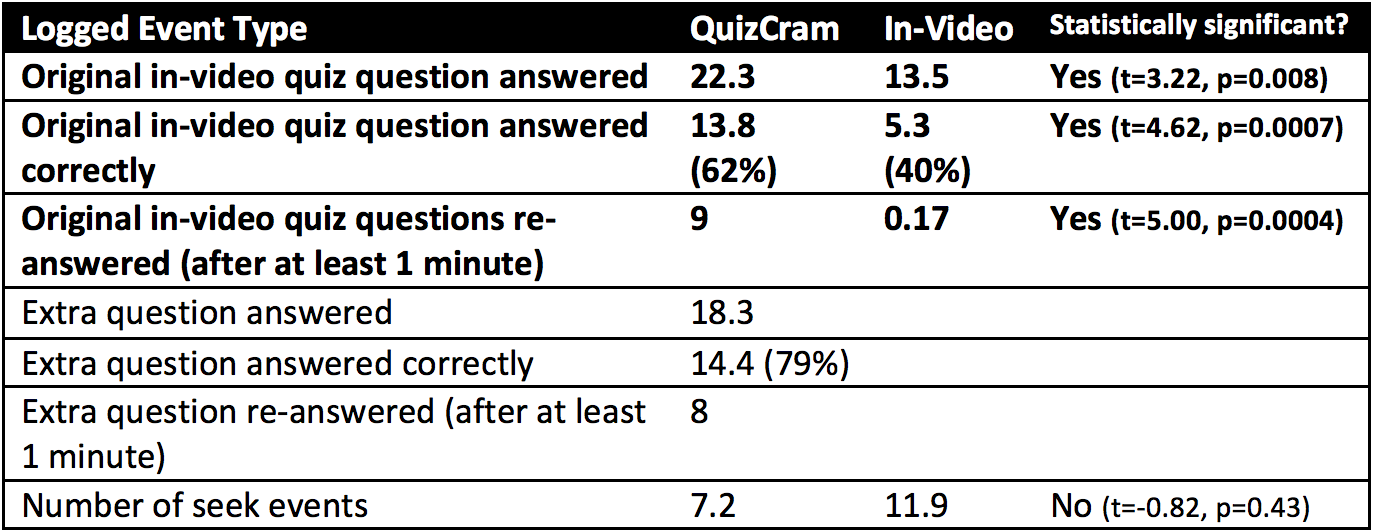}
\caption{Average number of events logged per user in each condition}
\label{fig:event-logs}
\end{figure}

\subsection{Analysis of Users' Video Interaction Logs}

% Users practiced answering questions more times when using QuizCram, as shown in \autoref{fig:event-logs}. Users also reviewed questions more often when using QuizCram. This increase in practice and reviewing helps explain the increased scores on the in-video questions during the exams.

To compare how users interacted with the two tools, we logged the users' interactions as they studied the lectures, as shown in \autoref{fig:event-logs}.

% This is likely due to QuizCram's question-driven video navigation: 
We found that users practiced answering each question more times when using QuizCram. They also tended to answer questions correctly a higher percentage of the time, perhaps because they had been able to preview the question before watching the video. Users also reviewed previously-answered questions more often when using QuizCram. This increase in practice and reviewing helps explain the increased exam scores on the original in-video questions. %during the exams. % when using QuizCram.

% To compare how users were using the two systems, we logged usage data for each tool, as shown in \autoref{fig:event-logs}. We found that users practiced answering questions more times when using QuizCram. Users also reviewed questions more often when using QuizCram. This increase in practice and reviewing helps explain the increased scores on the in-video questions during the exams. % when using QuizCram.

% To compare how users were using the two systems, we logged the number times users answered questions or seeked with each tool, as shown in \autoref{fig:event-logs}. We found that users practiced answering questions more times when using QuizCram. Users also reviewed questions more often when using QuizCram (we defined ``reviewing'' as re-answering a question at least a minute after they first answered it). This increase in practice and reviewing helps explain the increased scores on the in-video questions during the exams.

Users seeked less on average when using QuizCram, which may partly be because they did not have to seek to and from in-video quizzes. However, this difference was not statistically significant.

\section{Conclusion}
% =============================================================================
We have presented QuizCram, a system that guides users' video viewing using questions. QuizCram aims to:

\begin{compactitem}
\item Encourage users to answer and review questions while they watch videos
\item Enable users to easily follow question-driven video navigation strategies (which we currently observe some users already using on Coursera)
\end{compactitem}

QuizCram breaks the video into segments associated with questions, and shows a focus question alongside the video. This question serves as an advance organizer that guides the user's attention towards the key points in the video. QuizCram also encourages reviewing based on questions: it displays a timeline of questions previously answered and their associated videos. It keeps track of users' progress through questions and videos, and suggests questions for users to review. Courses in the QuizCram format can be generated from existing videos with in-video quizzes. %, though it also has the flexibility to accommodate additional questions.

% Based on our observations that in-video quizzes play an important role in video navigation, we designed QuizCram, a system that uses questions to direct users' viewing of the material. It allows users to preview the question for the sections, and provides a timeline of questions and suggestions to encourage reviewing.

Our user study found that QuizCram increases retention of questions -- when the in-video questions were tested a day later, QuizCram users remembered them better than if they were presented as in-video quizzes. Users practiced answering and reviewing questions more when using QuizCram. % We also found that increasing the amount of questions presented with QuizCram results in users being able to recall the material tested by the additional questions better. %, even when answering based on recall not recognition.

Our user study has focused on a cramming scenario -- where the user is trying to memorize a small amount of material to prepare for an imminent exam. However, another potential use case for QuizCram-like systems is for long-term retention -- where the user is attempting to remember the content of multiple courses over multiple months. Given the success of spaced repetition systems in helping users' long-term retention of flashcards and vocabulary, we expect that having a system schedule quizzes that review course contents should similarly be helpful for helping users' long-term retention of course materials. Studying how question-driven lecture-reviewing systems can scale to entire courses and longer study periods is potential future work. % For example, we could potentially scale the idea of testing the user with a question to find not just whether the user needs to review an individual video segment, but . % Other future work in the context of multi-course % Although evaluation in the context of long-term retention is considerably harder than a lab study, we expect that having a system schedule reviews when learning MOOC lecture contents should be helpful in these contexts. % Indeed, given that Coursera's logs indicate that over the course of a 10-week MOOC, only 11\% of users will ever review a given lecture video, this suggests that even within MOOCs, a means of scheduling reviews is necessary.

We designed QuizCram to address the needs of users who wish to complete the MOOC and master the entire material. Hence, the system tests users' knowledge of video segments, and schedules reviews to ensure that users remember the material. That said, learning the complete course material is not the objective of many learners  -- many users are only interested in a subset of the material, and do not complete the rest of the course \cite{deconstructing} \cite{anderson2014engaging}. Although addressing the needs of users interested in only a subset of the material was not an objective of QuizCram, it is potential future work. % Nevertheless, we did find that (see supplemental material). We to address . There is already a large body of systems designed to address the second, which we . %There are some users who . Skimming through a large body of videos to find , which our system does not attempt to address. Although our system allows users -- we did not .

Current online courses rely on external problem sets and exams to test understanding of content in more depth than the in-video quizzes. However, many MOOC participants interact primarily with videos and do not take exams or do problem sets \cite{deconstructing} \cite{anderson2014engaging}. Thus, moving more of the course content out of problem sets and making the video more interactive and question-oriented provides a way to benefit these viewers without removing them from the scaffolding of videos. We believe that QuizCram is a logical step from in-video quizzes towards more interactive, question-driven study experiences.

\balance{}

\bibliographystyle{acm-sigchi}
\bibliography{quizcram}

\end{document}